\documentclass[12pt]{iopart}
\usepackage{epsfig,cite}  
\begin{document}


\title[Josephson Current Counterflow in Atomic BEC's]{Josephson Current Counterflow in
Atomic Bose-Einstein Condensates}


\author{E. Sakellari, N. P. Proukakis, M. Leadbeater, and C. S. Adams}

\address{Department of Physics, University of Durham, Durham DH1 3LE, United Kingdom}

\begin{abstract}

Josephson weak links in superconductors can be engineered such that the phase difference across the junction is
 $\pi$ ($\pi$-junction), leading to the reversal of
the current. The conditions under which the analogous effect of supercurrent
counterflow can be achieved in a double-well atomic Bose-Einstein condensate are investigated. 
It is shown that this effect is observable for condensates of up to a few thousand atoms, 
which are initially prepared in
an anti-symmetric `$\pi$-state' and subsequently subjected to a uniformly increasing  magnetic field
gradient. This effect is found to be only weakly-dependent on trap geometry, and can be
observed in both attractive and repulsive condensates.\\
\\
\vspace{7cm}

\bf Note: A substantially revised version of this manuscript
(with various figures removed, improved discussion and some additional material) is available
on cond-mat/0312396.
\end{abstract}



\maketitle

\section{Introduction}

The creation of superconducting \cite{SC_WeakLink} and superfluid \cite{SF_WeakLink} weak links has led to the
experimental observation of Josephson effects \cite{Josephson}, arising as a result of macroscopic quantum phase 
coherence. 
Josephson weak links are typically created by connecting two initially independent 
systems (superconductors / superfluids)
via a barrier with dimensions of the order of the system healing length.
Such junctions lead
to a variety of interesting phenomena \cite{barone}, 
such as dc- and ac-Josephson effects, Shapiro resonances,
macrosopic quantum self-trapping and $\pi$-phase oscillations.
Observations in superconductors preceded those in superfluids, due to the much larger
healing lengths, thus enabling easier fabrication of weak links.
Evidence for Josephson-like effects has been observed in
$^{4}$He weak links \cite{Sukhatme}, and unequivocally demonstrated 
for weakly-coupled $^{3}$He systems \cite{avenel}.
The recent achievement of dilute trapped atomic Bose-Einstein condensation (BEC)
\cite{BEC}  gives rise to a new system for studying Josephson effects.  
In particular, such systems enable the investigation of
 dynamical regimes not easily accessible with other superconducting or superfluid systems. 
The simplest atomic Josephson junction can be realized by a condensate confined in a double-well potential.
To allow control of the tunnelling rate, such a system can be constructed by raising a barrier
within a harmonic trap containing an atomic condensate; 
 this can be achieved by the application of a far-off-resonant blue-detuned laser beam, which induces
a repulsive gaussian barrier \cite{MIT_DW}.
Atomic interferometry based on such a set-up was recently reported \cite{MIT_Interferometer}.
Alternatively, a condensate can be created directly in  a magnetic double-well structure \cite{Walraven}.
Remarkable experimental progress has led to the creation of atomic BEC Josephson junction arrays, in which
the harmonically trapped atoms are additionally confined by an optical lattice potential,
generated by far-detuned laser beams. Phase coherence in different wells was observed by
interference experiments of condensates released from the lattice \cite{Kasevich}.
In addition, Josephson effects \cite{Inguscio} and the control of tunnelling rate has been 
demonstrated \cite{Mott,NIST}. Although experiments (and theoretical analysis) of such systems are
well underway, deeper insight into the diverse range of Josephson phenomena can be obtained
by looking at the simplest single junction, double-well system.
This system has already received considerable theoretical attention, with treatments
based on a two-state approximation 
\cite{Two_State_0,Two_State_1,Two_State_2,Two_State_3,Two_State_4,Two_State_5,Janne}, 
zero temperature mean field theory 
\cite{MF_1,MF_2,fant,Two_State_6,MF_3,sak},
 quantum phase models \cite{Phase,Exact_Phase} and instanton methods \cite{Instanton}.

This paper focuses on a double-well atomic BEC, and investigates the conditions under which
the Josephson current can be engineered to flow in a direction opposite to that minimizing the 
potential energy. This phenomenon bears close analogies to superconducting $\pi$-junctions \cite{Pi_Junction}, in which
a macroscopic
phase difference  $\phi=\pi$ is maintained across the superconducting weak link.
Such behaviour has been observed in a variety of systems, as a consequence of different microscopic mechanisms.
For example, $\pi$-junctions in ceramic superconductors have their origin in the symmetry of the order
parameter \cite{Order}, with their experimental detection being central to 
the understanding 
of high-$T_{c}$ superconductivity.
$\pi$-junctions can also be created in ferromagnetic weak links \cite{Ryazanov}, or by magnetic impurities \cite{Magnetic}.
Recent interest has focused on the creation of controllable $\pi$-junctions in superconducting/normal-metal/superconducting
links, in which the current direction can be reversed by the application of suitably large voltage across the link \cite{Reversal}.
The reversal originates from the fact that the addition of an extra phase factor $\pi$ is equivalent to reversing the sign of the current $I_{c}$,
since the superflow current obeys $I=I_{c} \sin (\phi)$.
A controllable $\pi$-SQUID (Superconducting Quantum Interference Device) has been demonstated \cite{SQUID}, and it is clear that
the manipulation of multiple such $\pi$-junctions will be important in the domain of quantum electronics.
For example, an array of alternating $0-\pi$ junctions allows the spontaneous generation of half-integer flux quanta.
Such a circuit of multiple successive $0$-$\pi$ junctions has been recently created between thin films of
high-$T_{c}$ and low-$T_{c}$ superconductors \cite{Fluxons}, generating a one-dimensional array of Josephson contacts with alternating signs
of current.
The superfluid analogue of a supeconducting $\pi$-junction is a metastasble 
$\pi$-state. This was recently observed
in $^{3}$He weak links \cite{Pi_State} upon exciting the system by an
oscillating driving force.
Atomic BEC junctions behave similarly to those of $^{3}$He. For example, 
in the usual manner of considering mechanical analogs of Josephson 
junctions, superconducting Josephson junctions can be mapped onto a rigid pendulum, whereas atomic tunnel
junctions ($^{3}$He, BEC) map onto a non-rigid pendulum \cite{Pendulum}, thus exhibiting richer oscillation modes.
This model has been used to discuss so-called $\pi$ oscillations \cite{shen}
and their stability in atomic BEC's \cite{Pi_BEC}, while such states have also been shown
to arise within the framework of an exact quantum phase model \cite{Exact_Phase}.

In this paper, we investigate the circumstances under which one can reverse the direction of the
atomic current across a suitably-prepared condensate-condensate weak link, by the application of a linear potential gradient.
We find that a BEC confined in a double well configuration can, for small values of the potential gradient, move towards the 
higher potential well, a phenomenon henceforth referred to as Josephson counterflow. 
Although such Josephson counterflow bears close analogies to the behaviour observed
in superconducting weak links, we should point out that the counterflow discussed
in this paper is `global' (i.e. flow of entire quantum gas in opposite direction),
as opposed to `local' counterflow (i.e. across a single junction) in a superconducting 0-$\pi$ junction.

This paper discusses in detail the phenomenon of Josephson  counterflow for the lowest state exhibiting such behaviour, namely
the anti-symmetric first excited $\pi$-state, which is most
amenable to experimental observation. 
Atomic counterflow dynamics are investigated in terms of experimentally
 relevant parameters, such as
interaction strength, harmonic trap aspect ratio and gaussian barrier geometry.
Our analysis is based on numerical simulations of the Gross-Pitaevskii equation in three dimensions
and leads to the conclusion that there exists a realistic 
window of parameters in which atomic Josephson counterflow can be experimentally observed.
This effect is found to be only weakly dependent on the strength of the interactions and can, in prinicple,
be observed for both attractive and repulsive condensates.
One should note the distinction between the $\pi$-state considered in this paper which is a time-independent
solution, as opposed to the $\pi$ oscillations which arise as a result of a superposition of ground and first excited states.
In an experimental realization, it may be difficult to create a pure $\pi$-state, and the system may experience a
combination of counterflow and $\pi$-oscillations. 
In this paper we show that,
by careful initial state preparation, one can decouple the
timescales for these two effects, and even induce counterflow in a direction
opposite to that of $\pi$-oscillations, thus demonstrating the different origin of these two phenomena.
Note that Josephson counterflow has already been predicted in condensates
trapped in optical lattices, as a result of
 the renormalization of the mass in the lattice,
based on Bloch wave analysis \cite{levit}.
Such an interpretation is, however, not easily
transferable to the double-well system.

This paper is structured as follows.
Sec. 2 introduces our main formalism, outlining the behaviour of a double-well condensate.
Sec. 3 discusses the dynamics of atomic Josephson junctions, whereas the possibility of experimental 
observation of this phenomenon in current BEC set-ups is analysed in Sec. 4; finally, we conclude in Sec. 5.

\section{BEC in a double-well potential}

At low temperatures, the behaviour of a Bose-Einstein condensate is accurately described by a 
nonlinear Schr\"{o}dinger equation known as the
Gross-Pitaevskii (GP) equation.
Throughout this paper we work in dimensionless (harmonic oscillator) units, by
applying the following scalings:
space coordinates transform according to
$\mbox{\boldmath$r$}_{i}^{\prime}={a_{\perp}}^{-1}\mbox{\boldmath$r$}_{_{i}}$, time $t^{\prime}=\omega_{\perp} t$,  
condensate wavefunction 
$\psi^{\prime}\left(\mbox{\boldmath$r$}^{\prime},t^{\prime}\right)=\sqrt{a_{\perp}^3}\psi\left( \mbox{\boldmath$r$},t\right)$  and
 energy 
$E^{\prime}=\left(\hbar\omega_{\perp}\right)^{-1}E$.
Here $a_{\perp}=\sqrt{\hbar/m\omega_{\perp}}$ is the harmonic oscillator length
 in the transverse direction(s), where $\omega_{\perp}$ 
the corresponding trapping frequency.
We thus obtain the following dimensionless GP equation (primes henceforth neglected for convenience)
describing the evolution of the condensate wavefunction (normalised to unity)
\begin{eqnarray}
i \partial_{t}\psi \left( \mbox{\boldmath$r$},t\right) =  \left[ -{\frac{1}{2}} \nabla^{2}
+V\left( \mbox{\boldmath$r$}\right)+\tilde{g}|\psi(\mbox{\boldmath$r$},t)|^{2}\right] \psi (\mbox{\boldmath$r$},t)~.
\label{eq:GP3dhou}
\end{eqnarray} 
The atom-atom interaction is parametrized by $\tilde{g}=g/(a_{\perp}^3 \hbar \omega_\perp)$, where
 $g={\cal N}(4\pi\hbar^2a/m)$ is the usual three-dimensional scattering amplitude,
defined in terms of the {\it s}-wave scattering length $a$, and
 ${\cal N}$ is the total number of atoms (mass $m$).
 $V\left({\bf r}\right)$ represents the total 
confining potential.  
Steady state solutions of the GP equations can be obtained by substituting 
$\psi\left(\mbox{\boldmath$r$},t\right)=e^{-i\mu t}\phi \left(\mbox{\boldmath$r$}\right)$.

\begin{figure}[hbt]
\centering
\includegraphics[width=12.0cm]{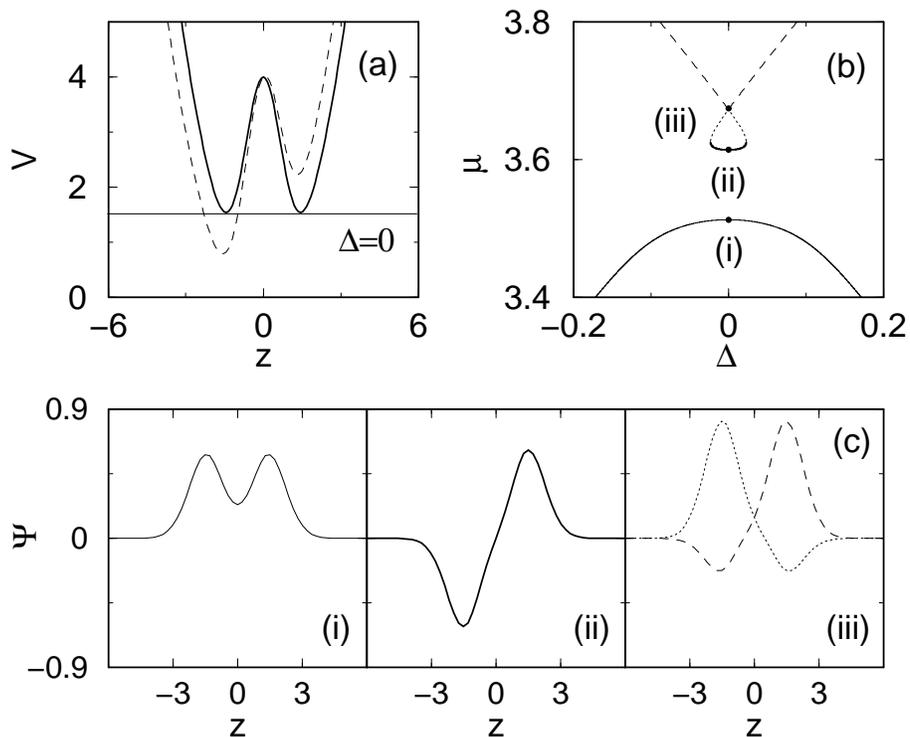}
\caption{
Double well potential with corresponding eigenenergies and eigenstates.
(a) Schematic geometry of the total  confining potential in the axial direction for a Gaussian barrier
(height $h=4 \hbar \omega_{\perp}$, waist $w=a_{\perp}$) located at the centre of the trap. 
Plotted are the symmetric  ($\Delta=0$, solid line) and asymmetric  
($\Delta =0.5 (\hbar \omega_{\perp} / a_{\perp})$, dashed line) case.
(b)-(c) Corresponding eigenenergies and eigenstates for the double-well as a function of
the  potential gradient $\Delta$:
(i) ground state (lower thin solid line), (ii) anti-symmetric first-excited $\pi$-state (thick solid line),
(iii) first excited state with unequal populations, having more population in left well (dotted),
or in right well (dashed).
Parameters used here are $\tilde{g}=\pi$ and spherical trap geometry ($\lambda=1$).
}
\end{figure}

In the double-well configuration, the
total confining potential is given by (see Fig. 1(a))
\begin{equation}
V\left({\bf r}\right)=\textstyle{1\over 2}\left[(x^2+y^2)+\lambda^2 z^2\right]+%
h\exp\left[-(z/w)^2\right]+\Delta z~.
\label{eq:conf3d}
\end{equation}
The first term describes a cylindrically symmetric harmonic trapping potential, with a trap aspect ratio
$\lambda=\omega_{z}/\omega_{\perp}$: the trap is spherical for $\lambda=1$,
`cigar-shaped' for $\lambda<1$  and `pancake-like'
 for $\lambda>1$.
The second term describes a gaussian potential of height $h$ generated by a blue detuned light sheet of
beam waist $w$ in the $z$ direction, located at $z=0$.
Finally, the contribution $\Delta z$ corresponds to a linear magnetic field gradient $\Delta$ pivoted at the
centre of the trap. The populations  of the two wells are equal for  $\Delta=0$ (symmetric case, solid line in Fig. 1(a)),
whereas  $\Delta > 0$ (dashed line in Fig. 1(a)) leads to a tilted potential, which induces  tunnelling.
The application of the magentic field gradient $\Delta>0$ additionally shifts the trap centre to the $z>0$ region.
However, this shift is negligible for the parameters studied throughout this work, and will be henceforth ignored.

The eigenenergy curves of the double-well condensate are calculated 
 by numerical solution of the time-independent 
GP equation, as discussed in more detail in our preceeding paper \cite{sak}.
Sufficiently large interactions lead to the appearance of a loop structure \cite{Two_State_5,Loop}.
The loop structure for the first excited state is shown in Fig. 1(b).
Corresponding wavefunctions for ground (solid) and first 
excited state 
(dashed) are shown in Fig. 1(c) for $\Delta=0$. The three eigenstates are
(i) a symmetric state with equal population in both wells (solid line), (ii) an anti-symmetric `$\pi$-state'
with equal population in both wells and a phase difference of $\pi$ across the trap centre, 
and (iii) two higher energy `self-trapped' 
states with most of the population in either the left (dashed) or the right (dotted) well. 

In this paper, we are mostly interested in the behaviour of the $\pi$-state. We will show that, if
the system is prepared in the $\pi$ state, the subsequent temporal evolution of the system
can exhibit Josephson counterflow.

\section{Josephson Counterflow Dynamics}

\begin{figure}[b]
\centering
\includegraphics[width=5.0cm,height=6.95cm]{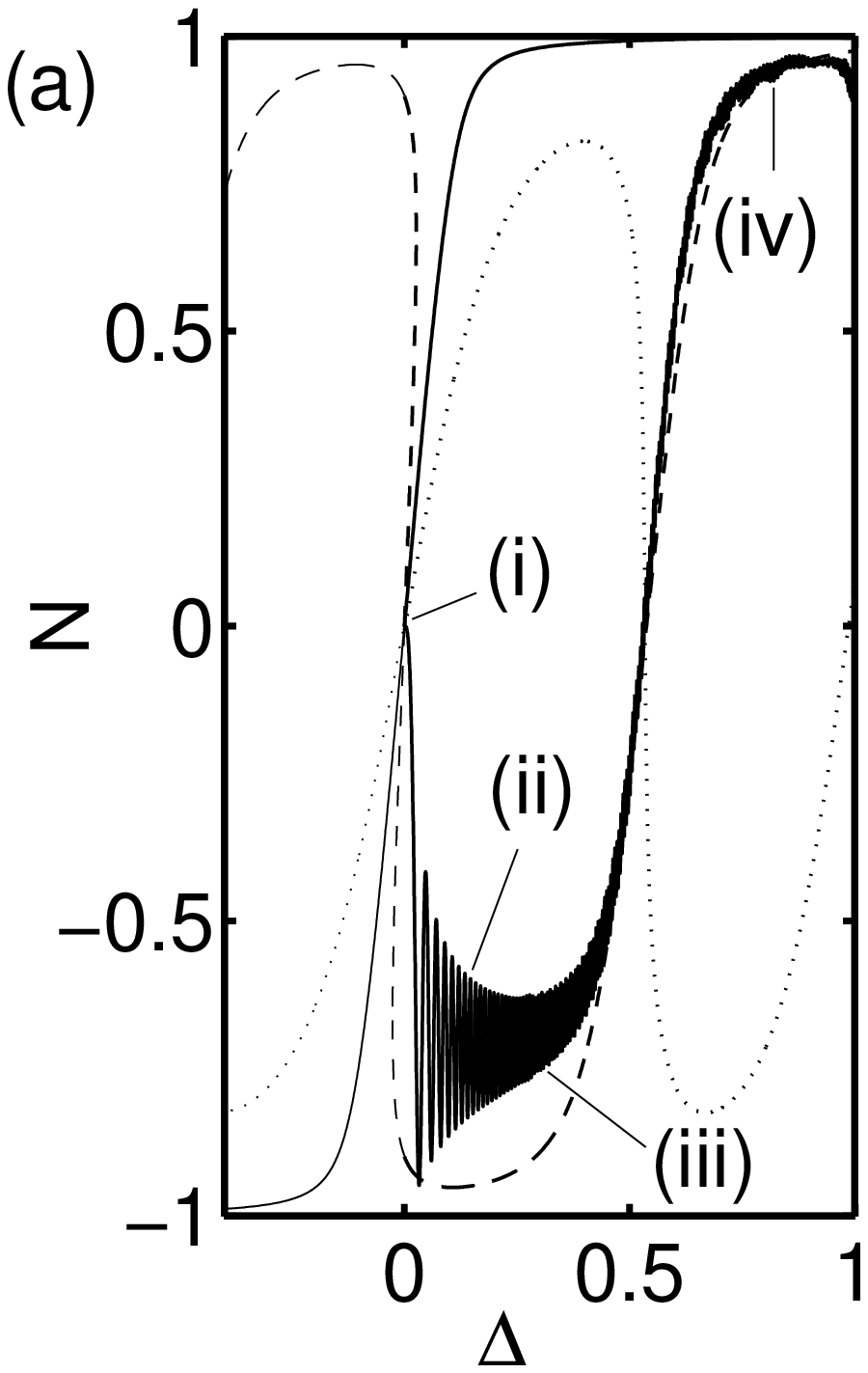}
\includegraphics[width=6.7cm]{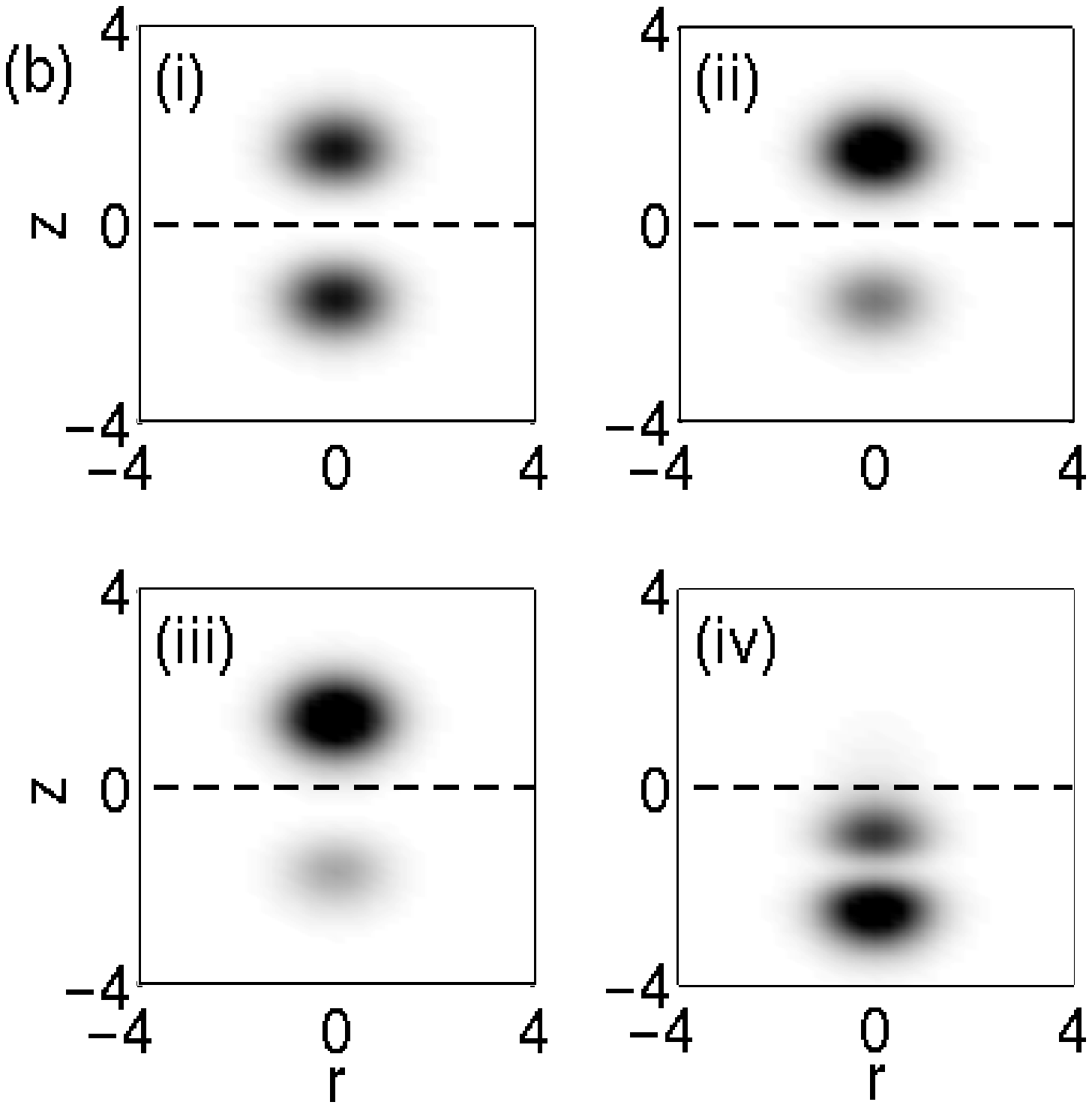}
\caption{
(a) Evolution of fractional population difference $N$ as a function of  potential gradient $\Delta$, 
for a system initially prepared in a $\pi$-state. Here  $h=4 \hbar \omega_{\perp}$, $\lambda=1$, $\tilde{g}=\pi$ and
 the potential gradient $\Delta=Rt$ increases 
at constant rate $R=10^{-3}  (\hbar \omega_{\perp}^{2} /a_{\perp})$.
The corresponding population differences for the ground state (thin solid), first excited (dashed) and second excited state (dotted) eigenstates
are also shown.
(b) Snapshots of the evolution of the density distribution for case (a)
when  (i) $\Delta= 0$, 
(ii) $0.1 $, (iii) $0.3  $
 and (iv) $0.8$ in units of $ (\hbar \omega_{\perp}/a_{\perp})$.
The population of both wells is initially
equal ($\Delta=0$). As the gradient is 
increased in (ii), (iii), population starts being transferred towards the right ($z>0$), upper well. Increasing the asymmetry beyond a threshold value 
leads the population to be once again transferred to the left ($z<0$), lower well. Eventually, (iv),  a transition 
to the second excited state occurs (d) (see movie).
} 

\end{figure}

To demonstrate counterflow, we study  Josephson dynamics of the $\pi$-state 
under a potential gradient $\Delta >0$, which is linearly increasing at rate $R$ (i.e. $\Delta=Rt$),
such that the right well lies higher than the left well (dashed line in Fig. 1(a)).
On the basis of the usual Josephson relations, one might naively expect the atomic current to flow
towards the region of lower potential, i.e. towards the left well.
Instead, we observe that, for suitable parameters (see later), superflow can occur from the 
lower potential energy well (left) to the higher potential energy one (right).
This is a direct consequence of the $\pi$ phase difference of the antisymmetric initial wavefunction,
and does not occur for a system in the symmetric ground state (for which flow
always occurs towards regions of lower potential energy - see, e.g. \cite{sak}).
Our study focuses on the dynamics of the fractional relative population, $N=(N_L-N_R)/(N_L+N_R)$,  between the two wells, as opposed
to the current through the junction (which is the derivative of $N$).
The dependence of $N$  on potential gradient $\Delta$
 is shown for a spherical trap in Fig. 2(a), for a system initially prepared in the $\pi$-state.
In such a state, the population of both wells is initially
equal ($\Delta=0$). As the gradient is 
increased, population starts being transferred towards the right, upper well, such that we observe 
Josephson counterflow to regions of higher potential energy.
However, increasing the gradient beyond a threshold value 
leads to suppression of this effect, with the population once again transferred to the left, lower well.
 Eventually, the perturbation due to the applied potential
becomes so pronounced, that  a transition 
to the second excited state occurs \cite{sak}.
Characteristic density snapshots of this evolution are shown in Fig. 2(b).
The initial counterflow dynamics can be understood  by means of lowest order perturbation theory. However, such a simple
picture no longer gives an accurate description when the population difference becomes large. The two-state model
\cite{Two_State_0,Two_State_1,Two_State_2,Two_State_3,Two_State_4,Two_State_5} 
also reproduces the initial counterflow dynamics.
However, since this model contains no mechanism for removing the system from the counterflow state,
the two-state model predicts
equilibration in a macroscopically quantum trapped state with larger population in the higher well. 
This inadequacy of the two-state model is based on
the fact that it does not take higher lying modes into consideration \cite{sak}.

\begin{figure}[t]
\centering
\includegraphics[width=14.0cm]{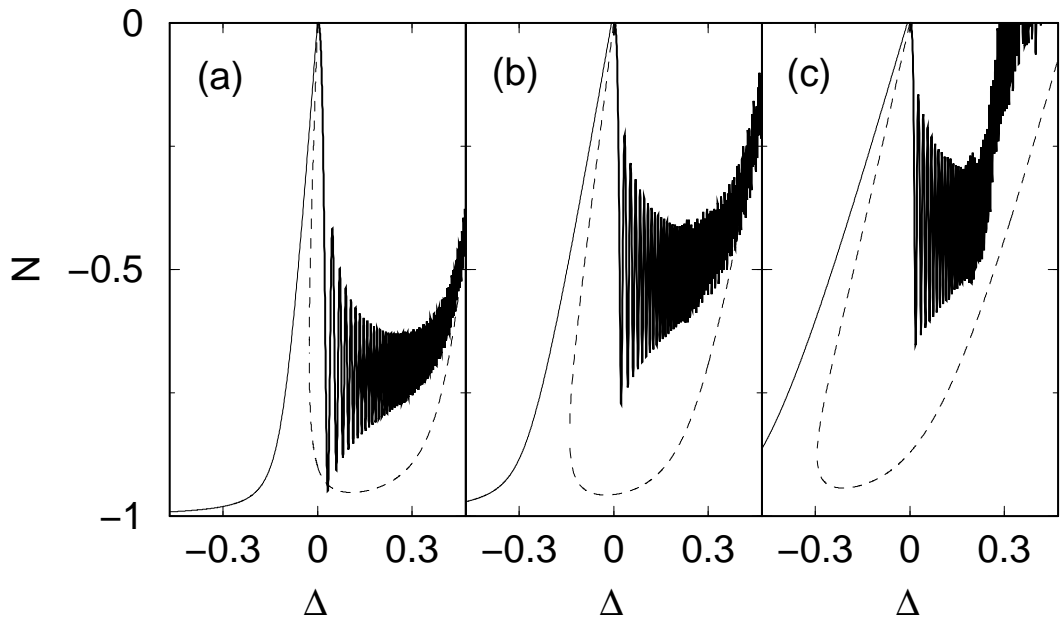}
\caption{Evolution of fractional population difference $N$ as a function of potential gradient $\Delta$ for
identical trap configurations ($\lambda=1$, $h=4 \hbar \omega_{\perp}$ and $R= 10^{-3} (\hbar \omega_{\perp}^{2}/a_{\perp})$) and 
increasing nonlinearity (a) $\tilde{g}=\pi$, (b) $4\pi$ and (c) $10\pi$.
Solid bold (dashed) lines indicate corresponding eigenstate populations for the ground (first excited)
state.}
\end{figure}  

Increasing the nonlinearity 
causes a reduction in the amount of
initial counterflow, and thus tends to inhibit the experimental observation.
Fig. 3 illustrates the reduction in counterflow due to increased nonlinearity for fixed trap geometry, with
$\tilde{g}$ increasing by a factor of $10$ from (a) to (c).
This would appear to restrict the observation of the phenomenon to moderate nonlinearities (see next section for experimental estimates).

\begin{figure}[hbt]
\centering
\includegraphics[width=14.0cm]{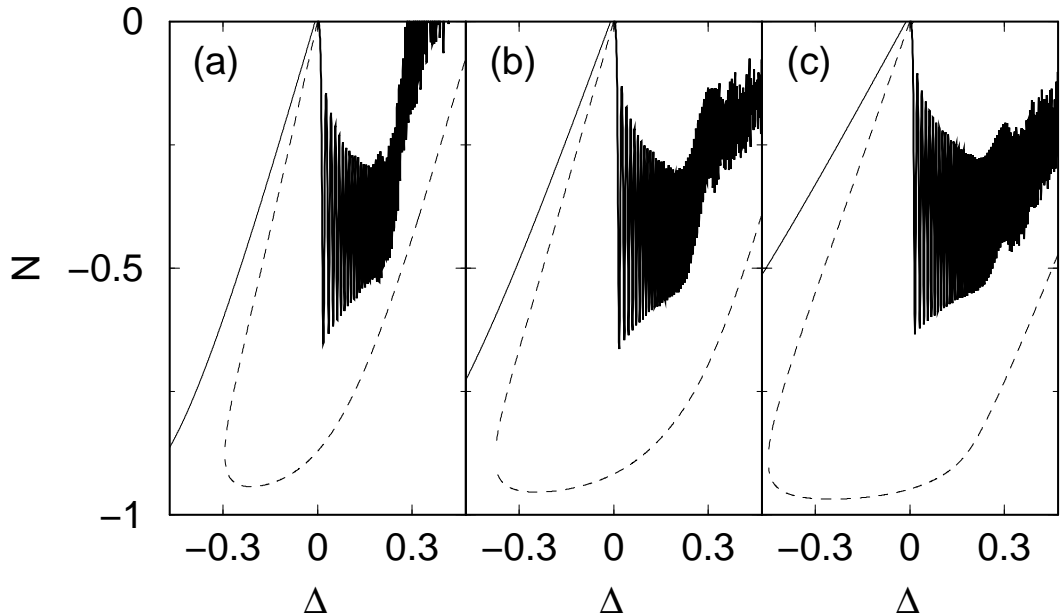}
\caption{
Dependence of fractional population difference dynamics on trap geometry.
Plotted is the evolution of $N$ as a function of potential gradient $\Delta$ for $R=10^{-3} (\hbar \omega_{\perp}^{2}/a_{\perp})$ (black), 
$\tilde{g}=10\pi$ and different trap aspect ratios: (a) cylindrically-symmetric trap $\lambda=1$ ($h=4 \hbar \omega_{\perp}$)
(same as Fig. 3(c)), (b)-(c) `pancake'-like traps with $\lambda=\sqrt2$ ($h=6 \hbar \omega_{\perp})$
 and (c) $\lambda=\sqrt8$ ($h=15 \hbar \omega_{\perp})$, respectively.
Note that larger barrier heights have been used with increasing aspect ratio, such that the density minimum at the trap centre is roughly
constant from (a)-(c).
}
\end{figure}

It is thus natural to ask if other factors (e.g. modifying initial trap aspect ratio, or changing barrier height $h$) will have the
opposite effect on the amount of counterflow, hence enabling observation of Josephson counterflow
even for large nonlinearities.
For example, tunnelling has been predicted to be enhanced for `pancake' traps ($\lambda > 1$) \cite{MF_3}. 
Indeed, for weak nonlinearities, such traps lead to a slightly increased counterflow amplitude.
Furthermore, such geometries feature enhanced energy splitting between ground
and first excited state, thus making them more
robust to coupling due to external (e.g. thermal\cite{Two_State_3,Two_State_4,Sols_NJ,Janne}) perturbations.
However, the reduction of the amplitude of counterflow due to increased nonlinearities tends to largely
suppress this geometry dependence, as shown in Fig. 4 for trap aspect ratios in the range  1 to $\sqrt{8}$.
In plotting this figure, the barrier height $h$ has been increased for larger $\lambda$, such
that the peak density in the trap centre remains essentially unchanged.

\begin{figure}[hbt]
\centering
\includegraphics[width=12.0cm]{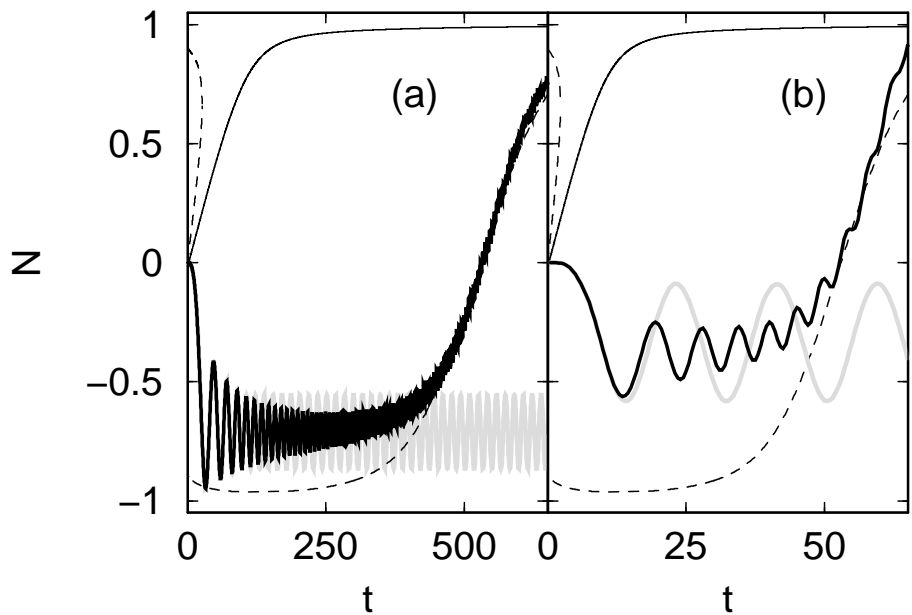}
\caption{
Evolution of fractional population difference $N$ as a function of time (thick black lines) for
different rates $R$ of increase of the potential gradient: (a) $R= 10^{-3}(\hbar \omega_{\perp}^{2}/a_{\perp}) $  and
 (b) $R= 10^{-2}(\hbar \omega_{\perp}^{2}/a_{\perp}) $.
Grey lines plot corresponding evolution of the population difference 
for the case when the potential gradient is held constant after a time $t=$ (a) $100 \omega_{\perp}^{-1}$ and (b) $10 \omega_{\perp}^{-1}$
Other parameters used, as in Fig. 3(a).}
\end{figure}

We should further comment on the extent to which our above findings depend on
 the rate $R$  with which the linear potential gradient $\Delta=Rt$ is applied.
The thick solid lines in Fig. 5 correspond to the evolution of the fractional population difference as a function of time.
Fig. 5(a) shows the dependence for  $R=10^{-3}  (\hbar \omega_{\perp}^{2}/a_{\perp})$ (as used in all earlier figures),
whereas Fig. 5(b) shows the corresponding behaviour when the gradient is increased at a rate ten times faster than (a).
One observes the following effects: firstly, the amount of maximum (initial) counterflow is significantly reduced
(roughly by a factor of 2) by increasing the rate $R$ by a factor of 10. Secondly, counterflow can only be observed for much
shorter times (roughly reduced by a factor of 10).

A final question of interest is what would happen to the population difference if the applied linear potential is ramped
 up to a particular value and subsequently kept constant.
The most striking behaviour will occur when the gradient is kept constant at the point of maximum population difference,
as indicated by the grey lines in Fig. 5.
We see that in this case, the population remains trapped in the right upper well, i.e. macroscopic quantum
self-trapping occurs to a state with higher potential energy. In this regime, where the gradient does not exceed the value
at which the flow is reversed, the two-state model predicts the behaviour correctly.

\section{Experimental Considerations}

Having demonstrated the existence of Josephson counterflow for a $\pi$-state initial wavefunction, we now discuss the
feasibility of such observation in atomic BEC experiments. Firstly, we need to discuss how such states 
with a node in the wavefunction and odd parity behaviour
can be generated.
Although not necessarily the most efficient method, here we consider phase imprinting \cite{Phase_Imprinting}.
Starting from the condensate ground state in a harmonic trap, 
population can be  transferred to the first excited state
by applying a light-induced potential of the form
\begin{equation}
V_{r}\left(z,t\right)=\alpha \sin \left(\pi t/\tau_{0}\right)\tanh \left(z\right)
\end{equation}
for ($t < \tau_{0}$),
where $\alpha$ and  $\tau_{0}$ are constants which we vary. 
This is equivalent to applying a $\pi$ phase shift to the system.
At $t=\tau_{0}$, the 
potential $V_{r}$ is suddenly switched off, and the  
system exhibits Rabi oscillations of variable amplitudes and frequencies between the
initial (ground) state and the first excited state.
Fig. 6 shows fractional occupations ((a), (c)) and corresponding population differences ((b), (d)) for
two different initial state preparation cycles.
The fractional occupation of state $i$, denoting here ground (upper, thick solid) or first excited state (thin solid line), as a function
of time, is given by
\begin{equation}
P\left(t\right)=\left|\left\langle \psi _{i}(z,t=0) |  \psi \left(z,t\right)\right\rangle\right|^2~.
\label{eq:prob}
\end{equation}
This quantity measures excited population in any combination of the first excited states, and not only the population in the
desired $\pi$-state. The subsequent Rabi oscillations are shown in Fig. 6(a).
The corresponding atom number population difference between the two wells
 features $\pi$-oscillations (Fig. 6(b)) \cite{Pendulum,shen},
in which there is almost complete exchange of atoms between the two wells
(i.e. essentially population exchange between the two macroscopically quantum self-trapped states of Fig. 1(c) (iii)). 
This phenomenon acts independently of the linear
magnetic field gradient and tends to obscure the effect of Josephson counterflow.
To demonstrate counterflow in its purest form, we thus consider the case (Fig. 6(c)) when the amplitude of such $\pi$-oscillations is suppressed
(Fig. 6(d)), 
such that  the majority of the population
remains in the left, lower well at all times.
Note that, in all cases studied here, the coupling with higher lying states is found to be negligible.

\begin{figure}[hbt]
\centering
\includegraphics[width=12.0cm]{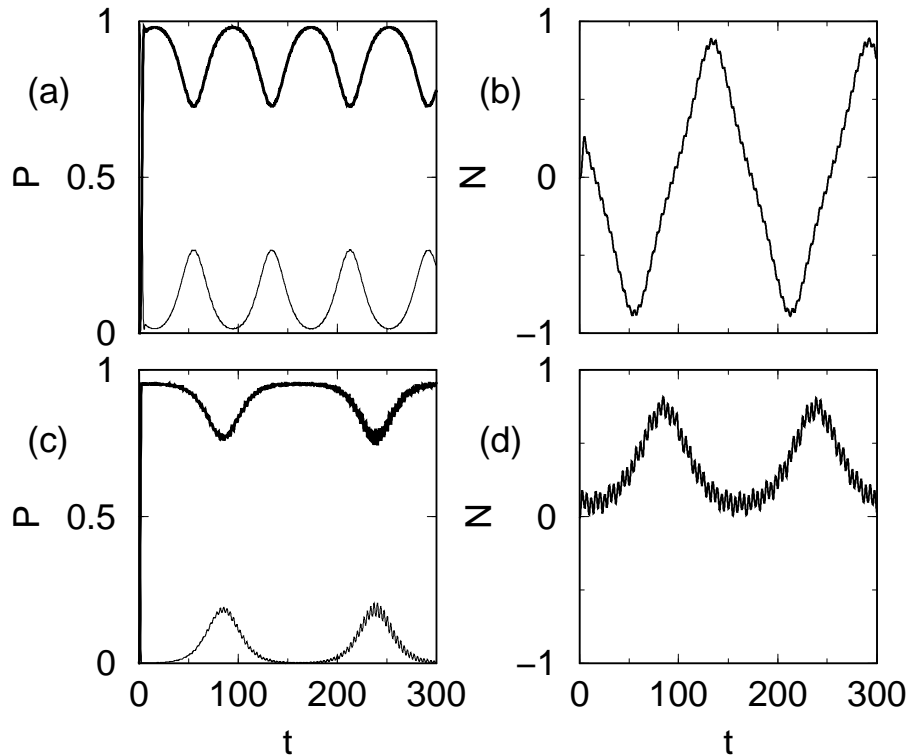}
\caption{
(a) Initial excited state preparation cycle and subsequent free evolution featuring Rabi oscillations
($\tau_{0}=6.5 \omega_{\perp}^{-1}$, $\alpha= 0.5 \hbar \omega_{\perp}$).
(b) Corresponding fractional population differences, exhibiting
 $\pi$-oscillations.
(c)-(d) Corresponding plots for different initial state preparation cycle  ($\tau_{0}=3  \omega_{\perp}^{-1}$,
$\alpha= \hbar \omega_{\perp}$).
 In (d), the $\pi$-oscillations have been suppressed, 
with macroscopic quantum self-trapping occuring in the left, lower well. 
Other parameters as in Fig. 3(a).}
\end{figure}

The importance of counterflow can be stressed, by first showing  typical fractional  population difference dynamics for a 
 double well condensate in its {\em ground} state \cite{sak}.
 Application of the external potential at $t=0$ creates a potential difference between the two wells,
with flow occuring towards the lower well (located on the left for $\Delta >0$). For the relatively slow rates
of increase of the potential gradient studied here, the system follows the ground eigenstate almost adiabatically (Fig. 7(b)).
In contrast, Figs. 7(c)-(d) show typical counterflow dynamics induced by the potential gradient $\Delta \neq 0$.
This is based on the initial state
 preparation  of Fig. 6(c)-(d) for $t \leq \tau_{0}$ and subsequent
free evolution for $\tau_{0} < t < \tau_{1}$ (as shown in  Fig. 7(a)), with the potential gradient $\Delta$ applied at $t=\tau_{1}$.
The evolution of the fractional population difference during this entire process is shown respectively
by the solid lines in Fig. 7(c)-(d) for (c) $\tau_{1}=10 \omega_{\perp}^{-1}$ 
and (d) $\tau_{1}= 85 \omega_{\perp}^{-1}$. 
The corresponding times when the potential gradient is applied are indicated by open circles in Fig. 7(a),(c)-(d).
The effect of Josephson counterflow manifests itself clearly in that the population starts being trasferred to the right, higher potential energy well.
Note that this is a direct consequence 
of the imposed magnetic field gradient (black curve), and that counterflow here occurs 
in a direction opposite to, and for larger amplitude than, the suppressed $\pi$-oscillations (grey lines).
This indicates clearly the
distinction between counterflow and $\pi$-oscillations.

\begin{figure}[hbt]
\centering
\includegraphics[width=12.0cm]{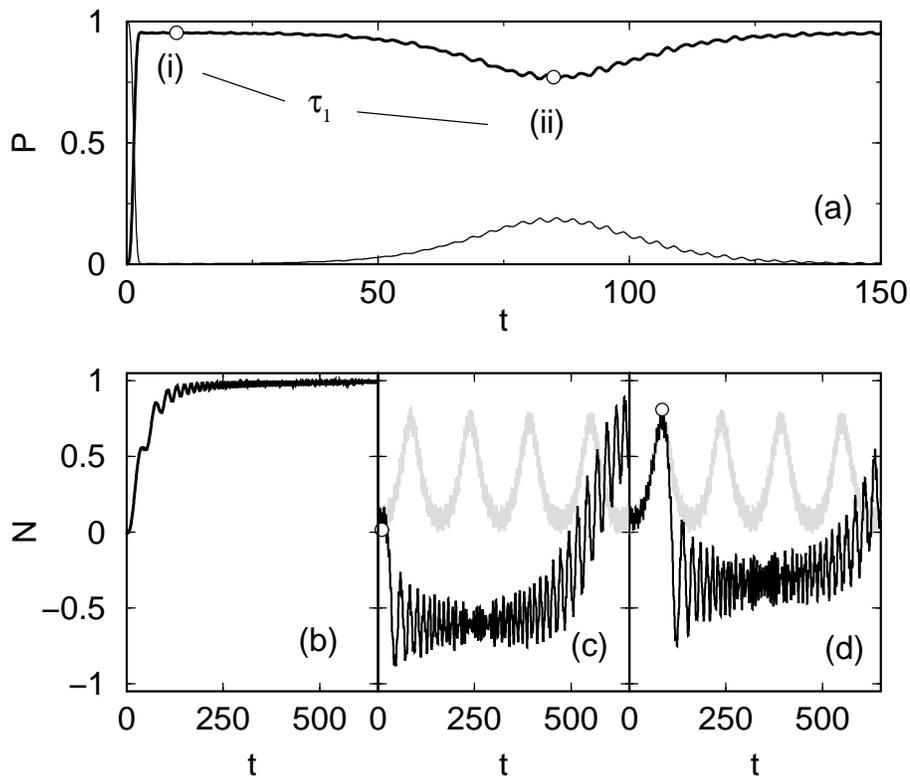}
\caption{
(a) Optimized $\pi$-state preparation cycle and subsequent free evolution featuring Rabi oscillations. (i) Maximum and (ii) minimum achievable
efficiency of population transfer to first excited state ($\tau_{0} = 3 \omega_{\perp}^{-1}$, $\alpha=\hbar \omega_{\perp}$).
(b)-(d) Evolution of fractional population difference $N$ as a function of potential gradient $\Delta$ for different initial states:
(b) Ground state condensate exhibiting usual Josephson flow towards lower (left) well, 
(c) Josephson counterflow (black) after efficient $\pi$-state preparation, induced by application of a potential gradient $\Delta=R(t-\tau_{1})$
for $\tau_{1}=10 \omega_{\perp}^{-1}$ (point (i) in Fig. (a)). Grey: corresponding evolution in the absence of the potential gradient, 
exhibiting (suppressed) $\pi$-oscillations.
(d) Same as (c), but with $\tau_{1}=85 \omega_{\perp}^{-1}$ (point (ii) in Fig. (a)).
Other parameters as in Fig. 3(a).}
\end{figure}  

We now look into typical experimental parameters which allow for demonstration of counterflow.
In particular, we should investigate whether this effect is observable for an experimentally realistic number
of atoms in the double-well condensate. The number of atoms is given by
\begin{equation}
{\cal N}={\frac{\tilde{g}}{4\pi}}{\frac{a_{\perp}}{a}}={\frac{\tilde{g}}{4\pi a}}\sqrt{{\frac{\hbar}{m\omega_{\perp}}}}~.
\label{eq:numberpart}
\end{equation}
It follows that, for given dimensionless nonlinearity $\tilde{g}$, large condensate atom number can be obtained
for light, weakly-interacting, transversally weakly-confined systems.
Note also that the total atom number is independent of the trap aspect ratio, as this cancels out for fixed $\tilde{g}$.
Hence, for this effect to be observed clearly, with a large number of atoms, one should preferably choose atoms with
a small value of $a \sqrt{m}$. This will hence yield large atom numbers for $^{7}$Li and $^{23}$Na, with the
corresponding numbers for $^{87}$Rb, $^{85}$Rb considerably smaller. Nonetheless, this effect should be observable for all above
species in experiments with resolution ability of detecting more than 1000 atoms. For example, taking $\tilde{g}=4 \pi$
as in Fig. 3(b) and $\omega_{\perp} = 2 \pi \times 5$ Hz, we find the following atom numbers in the double well:
${\cal N}=$  3300 ($^{23}$Na) and 1000 ($^{87}$Rb). This number could be enhanced by a factor
of 5 when using the nonlinearity of Fig. 3(c) and $\omega_{\perp} \sim 2 \pi \times 1 $ Hz, whereas further enhancement by a factor
of 10 is possible by tuning around a Feshbach resonance (e.g. $^{23}$Na, $^{85}$Rb, 
$^{133}$Cs \cite{Fesh}). 
In the case of $^{7}$Li, this effect should be observable in a very clean
manner, since in this case the constraint is placed on the maximum number of atoms which can be condensed such that it
does not exceed the critical number leading to collapse \cite{Rice_Collapse}. In the case of $^{85}$Rb,
the number of atoms needed to observe Josephson counterflow is not likely to exceed the critical value for collapse. 
The effect of Josephson tunnelling on collapse will be investigated in subsequent work \cite{sak_2}.

The possibility to demonstrate this effect experimentally will also depend on the timescales required for its observation. 
We consider the case of an applied magnetic field gradient $R=(10^{-3}-10^{-2})(\hbar \omega_{\perp}/a_{\perp})$ and a typical transverse
trapping frequency 
$\omega_{\perp} =2 \pi \times (5-100)$Hz. This translates into a magnetic field gradient inducing a Zeeman shift in the range
(1MHz$-$1GHz)/cm, and a  dimensionless timescale of $\omega_{\perp}^{-1}=(32-1.6)$ms. Hence,
for the illustrative parameters chosen here, efficient preparation of the $\pi$ state requires a time $\tau_{0} \sim (300-150)$ms.
Observation of Josephson counterflow requires monitoring the population difference for at least a further time 
$t_{\rm exp} \sim$ (1.5 s$-$75 ms).
One notices two competing effects here: For fixed, reasonably small, nonlinearity ($\tilde{g} < 10 \pi$), such that the effect can be
clearly observed, one needs weak transverse confinement   $\omega_{\perp}$ in order to obtain a reasonable number of atoms which can
be imaged easily. Contrary to this, small  $\omega_{\perp}$ imply long timescales, such that the observation of this effect becomes limited
by other factors (e.g. thermal damping \cite{Two_State_3,Two_State_4,Sols_NJ}, 3-body recombination, etc.). 
The best conditions will hence depend on the details of 
each set-up, with reasonable parameters in the range $ \omega_{\perp} \sim 2 \pi \times (5-100)$Hz and
$\omega_{z} \sim 2 \pi \times (1-500)$Hz, leading to total experimental timescales of $t_{\rm exp} \sim$ 100 ms$-$2 s, and requiring an optimum
resolution of a few hundred to a few thousand atoms.

\section{Conclusions}

We have studied the Josephson dynamics of a condensate in a double-well potential in the
 presense of a magnetic 
field gradient, for a suitably prepared initial antisymmetric state
featuring equal populations in each well and a $\pi$-phase slip across the weak link.
Under appropriate conditions, 
the atomic current was shown to flow in the direction opposite to that of minimum potential
energy. This is the opposite behaviour to the `normal' Josephson flow occuring for a system
in its ground state.
This phenomenon, termed here Josephson counterflow,  bears close analogies to
(metastable) $\pi$-states observed in superfluid-$^{3}$He and (controllable) $\pi$-junctions in superconducting weak links.
We have discussed a range of typical parameters for which this effect could be observed  in atomic
Bose-Einstein condensates.
Observation of this effect in a `clean' manner requires reasonably light, weakly-interacting atoms under rather weak transverse confinement,
and techniques to populate the first excited state in a manner such that the $\pi$-oscillations are heavily suppressed.
Optimum choice of parameters is an interplay between good experimental resolution for detecting few hundred to few thousand atoms, and the
maximum observation time for which this effect is not affected by other dephasing processes.
It is important to stress that appearance of this effect is not dependent on sign, and only weakly-dependent on strength of the scattering length,
applying equally well to both attractive and repulsive Bose-Einstein condensates.
An alternative suitable candidate for observing counterflow might also be found
in the recently realized atom chips \cite{Atom_Chips}, which offer excellent control and experimental resolution.
Investigation in these low dimensional systems should enable observation of Jospehson counterflow,
 since we have found this effect
to be only weakly dependent on aspect ratio of the harmonic trap.
We believe that the experimental observation of this phenomenon will further strengthen the analogies between atomic BEC's
and other states exhibiting macroscopic phase coherence and controllable Josephson effects.

\ack We acknowledge funding from the U.K. EPSRC.



\vspace{2.0cm}
\begin{thebibliography}{}

\bibitem{SC_WeakLink} Likharev 1979 Rev. Mod. Phys. {\bf 51} 101


\bibitem{SF_WeakLink} Davis J C and Packard R E 2002 Rev. Mod. Phys. {\bf 74} 741


\bibitem{Josephson} B. D. Josephson, Phys. Rev. Lett. {\bf 1}, 251 (1962)

\bibitem{barone} Barone A and Paterno G \textit{Physics and Applications of the Josephson Effect} 
(Wiley, New York, 1982).

\bibitem{Sukhatme} Sukhatme K, Mukharsky Yu, Chul T and Pearson D 2001 Nature {\bf 411} 280

\bibitem{avenel} Avenel O and Varoquaux E 1988 Phys. Rev. Lett. {\bf 60} 416 \\
Pereverzev S V, Loshak A, Backhaus S, Davis J C and Packard R E 1997 Nature {\bf 388} 449 \\
Backhaus S, Pereverzev S V, Loshak A, Davis J C and Packard R E 1997 Science {\bf 278} 1435.

\bibitem{BEC} M. H. Anderson \textit{et al.} 1995 Science {\bf 269} 198 \\
 Davis K B et al 1995 Phys. Rev. Lett. {\bf 75} 3969 \\
 Bradley C C et al 1997 Phys. Rev. Lett. {\bf 75} 1687 \\
Bradley C C et al. 1997 Phys. Rev. Lett. {\bf 79} 1170 \\
Fried D G et al. 1998 Phys. Rev. Lett. {\bf 81} 3811 \\
Robert A et al. 2001 Science {\bf 292} 461 \\
Dos Santos F P et al. 2001 Phys. Rev. Lett. {\bf 86} 3459 \\
Yosuke T, Maki K, Komori K, Takano T, Honda K, Kumakura M, Yabuzaki T and Takahashi Y 2003 Phys. Rev. Lett. {\bf 91} 040404 

\bibitem{MIT_DW} Andrews M R, Townsend C G, Miesner H J, Durfee D S, Kurn D M and Ketterle W 1997 Science {\bf 275} 637

\bibitem{MIT_Interferometer} Shin Y, Leanhardt A E, Saba M, Pasquini T, Ketterle W and Pritchard D E 2003 Nature (subm.)

\bibitem{Walraven} Tiecke T G, Kemmann M, Buggle Ch, Shvarchuck I, von Klitzing W and 
Walraven J T M 2003 J. Opt. B:Quantum Semiclass. Opt. {\bf 5} S119

\bibitem{Kasevich} Anderson B P and Kasevich M A 1998 Science {\bf 282} 1686 \\
Orzel C, Tuchman A K, Fenselau M L, Yasuda M and Kasevich M A 2001 Science {\bf 291} 2386

\bibitem{Inguscio} Cataliotti F S, Burger S, Fort C, Maddaloni P, Minardi F, Trombettoni A, Smerzi A, 
and Inguscio M 2001 Science {\bf 293} 843



\bibitem{NIST} Denschlag J H, Simsarian J E, H\"{a}ffner H, ${\rm M^{\rm c}}$Kenzie C, Browaeys A, Cho D,
Helmerson K, Rolston S L and Phillips W D 2002 J. Phys. B: At. Mol. Opt. Phys. {\bf 35} 3095

\bibitem{Mott} Greiner M, Mandel O, Esslinger T, H\"{a}nsch T W and Bloch I 2002 Nature {\bf 415} 39


\bibitem{Two_State_0} Jack M W, Collett M J and Walls D F 1996 Phys. Rev. A {\bf 54} R4625

\bibitem{Two_State_1} Milburn G J, Corney J, Wright E M and Walls D F 1997 Phys. Rev. A {\bf 55} 4318 

\bibitem{Two_State_2} Smerzi A, Fantoni S, Giovanazzi S and Shenoy S R 1997 Phys. Rev. Lett. {\bf 79} 4950

\bibitem{Two_State_3} Zapata I, Sols F and Leggett A 1998 Phys. Rev. A {\bf 57} R28

\bibitem{Two_State_4} Javaneinen J and Ivanov M Yu 1999 Phys. Rev. A {\bf 60} 2351

\bibitem{Two_State_5} Wu B and Niu Q 2000 Phys. Rev. A {\bf 61} 023402

\bibitem{Janne} Ruostekoski J and Walls D F 1998 Phys. Rev. A {\bf 58} R50


\bibitem{MF_1} Williams J, Walser R, Cooper J, Cornell E and Holland M 1999 Phys. Rev. A {\bf 59}, R31

\bibitem{MF_2} Salasnich L, Parola A and Reatto L 1999 Phys. Rev. A {\bf 60} 4171

\bibitem{MF_3} Williams J 2001 Phys. Rev. A {\bf 64} 013610

\bibitem{fant} Giovanazzi S, Smerzi A and Fantoni S 2000 Phys. Rev. Lett. {\bf 84} 4521

\bibitem{Two_State_6} Menotti C, Anglin J R, Cirac J I and Zoller P 2001 Phys. Rev. A {\bf 63} 023601

\bibitem{sak} Sakellari E, Leadbeater M, Kylstra N J and Adams C S 2002 Phys. Rev. A {\bf 66} 033612 

\bibitem{Phase} Leggett A J and Sols F 1991 Found Phys {\bf 21} 353

\bibitem{Exact_Phase} Anglin J R, Drummond P and Smerzi A 2001 Phys. Rev. A {\bf 64} 063605

\bibitem{Instanton} Zhou Y, Zhai H, Lu R, Xu Z and Chang L 2003 Phys. Rev. A {\bf 67} 043606 \\
Li W-D, Zhang Yu and Liang J-Q 2003 Phys. Rev. A {\bf 67} 065601

\bibitem{Pi_Junction} Bulaevskii L N, Kuzii V V and Sobyanin A A 1977 JETP Lett. {\bf 25} 290 \\
Geshkenbeim V B, Larkin A I and Barone A 1987 Phys. Rev. B {\bf 36} 235

\bibitem{Order} van Harlingen D J 1995 Rev Mod Phys {\bf 67} 515

\bibitem{Ryazanov} Ryazanov V V, Oboznov V A, Rusanov A Yu,Veretennikov A V, Golubov A A and Aarts J 2001
Phys. Rev. Lett. {\bf 86} 2427

\bibitem{Magnetic} M\"{u}hge Th et al 1996 Phys. Rev. Lett. {\bf 77} 1857

\bibitem{Reversal}  Baselmans J J A, Morpurgo A F, van Wees B J and Klapwijk T M 1999 Nature {\bf 397} 43

\bibitem{SQUID} Baselmans J J A, van Wees B J and Klapwijk T M 2001 Appl. Phys. Lett. {\bf 79} 2940

\bibitem{Fluxons} Smilde H J H, Ariand, Blank D H A, Gerritsma G J, Hilgenkamp H and Rogalla H 2002
Phys. Rev. Lett. {\bf 88} 057004

\bibitem{Pi_State} Backhaus S, Pereverzev S, Simmonds R W, Loshak A, Davis J C and Packard R E 1998
Nature {\bf 392} 687

\bibitem{Pendulum} Marino I, Raghavan S, Fantoni S, Shenoy S R and Smerzi A 1999 Phys. Rev. A {\bf 60} 487
\bibitem{shen} Raghavan S, Smerzi A, Fantoni S, and Shenoy S R 1999 Phys. Rev. A {\bf 59} 620
\bibitem{Pi_BEC} Raghavan S, Smerzi A and Kenkre V M 1999 Phys. Rev. A {\bf 60} R1787

\bibitem{levit} H. Pu, L. O. Baksmaty, W. Zhang, N. P. Bigelow and P. Meystre, Phys. Rev. A {\bf 67}, 043605 (2003).

\bibitem{Loop} Wu B, Diener R B and Niu Q 2002 Phys. Rev. A {\bf 65} 025601 \\
Diakonov D, Jensen L M, Pethick C J and Smith H 2002 Phys. Rev. A {\bf 66} 013604 \\
Mueller E J 2002 Phys. Rev. A {\bf 66} 063603


\bibitem{Phase_Imprinting} Burger S et al. 1999 Phys. Rev. Lett. {\bf 83} 5198 \\
Denschlag J et al. 2000 Science {\bf 287} 97

\bibitem{Fesh} Inouye S, Andrews M R, Stenger J, Miesner H-J, Stamper-Kurn D M and Ketterle W 1998
Nature {\bf 392} 151 \\
Roberts J L, Claussen N R, Cornish S L and Wieman C E 2000 Phys. Rev. Lett. {\bf 85} 728 \\
Robert A et al. 2001 Science {\bf 292} 461 


\bibitem{Rice_Collapse} Bradley C C, Sackett C A and Hulet R G 1997 Phys. Rev. Lett. {\bf 78} 985

\bibitem{sak_2} Sakellari E, Proukakis N P and Adams C S, 2003 In Preparation 


\bibitem{Sols_NJ} Kohler S and Sols F 2003 New J. Phys. {\bf 5} 94

\bibitem{Atom_Chips} Ott H, Fortagh J, Schlotterbeck G, Grossmann A and Zimmermann C 2001
Phys. Rev. Lett. {\bf 87} 230401 \\
H\"{a}nsel W, Hommelhoff P, H\"{a}nsch T W and Reichel J 2001 Nature {\bf 413} 498 \\
Schneider S, Kasper A, vom Hagen Ch, Bartenstein M, Engeser B, Schumm T, Bar-Joseph I, Folman R, Feenstra L and Schiedmayer J
2003 Phys. Rev. A {\bf 67} 023612 \\
Horak P, Klappauf B G, Haase A, Folman R, Schmiedmayer J, Domokos P and Hinds E A 2003 Phys. Rev. A {\bf 67} 043806












\end{thebibliography}
\end{document}